\documentclass[usenatbib]{mn2e}

\usepackage{color}
\usepackage{graphicx}
\usepackage{epsf}
\usepackage{epsfig,graphicx,psfrag,amsfonts,amssymb}
\usepackage{amsmath}
\usepackage{dcolumn}
\usepackage{bm}

\begin{document} 
\title[]
{Constraining Thawing Quintessence}
\author[]
{Gaveshna Gupta$^{1}$\thanks{gaveshna.gupta@gmail.com }, Subhabrata Majumdar$^{2}$\thanks{subha@tifr.res.in},  Anjan A Sen$^1$\thanks{anjan.ctp@jmi.ac.in}\\
$^1$ Center for Theoretical Physics, Jamia Millia Islamia, New Delhi-110025, India\\
$^2$ Department of Theoretical Physics, Tata Institute For Fundamental Reserach, 1 Homi Bhabha Road,  Mumbai 400005, India}
\date{\today}
\maketitle

\begin{abstract}
We look at observational constraints on the  thawing class of scalar field models proposed to explain the late time acceleration of the universe. Using the recently introduced `Statefinder Hierarchy', we compare these thawing class of models with other widely studied dark energy (and modified gravity) models to check the underlying parameter degeneracies. We put constraints on the deviations of these thawing models from the canonical  $\Lambda$CDM model using a large class of observational data, e.g,  the Supernova Type Ia data, the BAO data, the CMB data and data from the measurements of the Hubble parameter using red-envelope galaxies. We also forecast constraints using a simulated dataset for the  future JDEM SNe  survey. Our study shows that, although with current data it is difficult to distinguish different thawing models from $\Lambda$CDM,  a future JDEM like mission would be able tell apart thawing models from $\Lambda$CDM for currently acceptable values of $\Omega_{m0}$. 
\end{abstract}


\section{Introduction}
One of the great mysteries of the universe today is the nature of dark energy which drives the late time cosmic acceleration. This has been confirmed by number of observational results including Type-Ia Supernovae \citep{Riess1998,Perlmutter1999,Tonry2003,Knop2003,Riess2004}, cosmic microwave background radiation (CMBR) (\ Komatsu et al.~(2011)) as well as the latest surveys of the large scale structure (\ Eisenstein et al.~(2005)). There has been wide range of contrasting proposals to explain this late time cosmic acceleration. Most popular of these is to include a dark candidate with a large negative pressure (also known as dark energy) in the energy budget of the universe (\ Bean et al.~(2005), \ Copeland et al.~(2006), \ Li et al.~(2011), \ Padmanabhan(2003), \ Peebles \& Ratra ~(2003), \ Sahni \& Starobinsky~(2000)).

The simplest candidate for this dark energy is a cosmological constant $\Lambda$. However, the well-known  fine tuning and coincidence problems render $\Lambda$ a rather unattractive proposal from a theoretical point of view.  Going beyond $\Lambda$, scalar field models with generic features are perhaps the simplest alternatives to a cosmological constant. A large class of scalar field models including quintessence (\ Ratra \& Peebles~(1988),\ Caldwell et al.~(1998),\ Liddle \& Scherrer~(1999),\ Steinhardt et al.~(1999)),
tachyon (\ Abramo \& Finelli~(2003),\ Bagla et al.~(2003),\ Aguirregabiria \& Lazkoz~(2004),\ Copeland et al.~(2005)) phantom (\ Caldwell~(2002)) and k-essence (\ Armendariz et al.~(2001),\ Scherrer~(2004), \ Sen~(2006), \ Vikman~(2005)) have been thoroughly investigated in recent years to explain the late time acceleration of the universe. The advantage of these scalar field models is that not only can they alleviate the fine tuning and coincidence problem, they can also mimick a cosmological constant at present epoch. These models have equation of state as a function of redshifts. This feature is important to distinguish these scalar field models from cosmological constant as large amount observational data from higher redshifts are coming in or expected to come in near future. 

The simplest scalar field is the one with linear potential having a canonical kinteic energy. In this case, the scalar field is frozen initially due to large Hubble damping and behaves like a cosmological constant with $w \sim -1$. As the universe expands and Hubble damping decreases, the field starts rolling and the equation of state starts deviating away from $w=-1$.
But because the potential has no minima, the model gives a collapsing universe in future which results in a finite history for our universe (\ Kratochvil et al.~(2004)). 

More complicated scalar field models can be divided in to two broad classes: the fast roll and slow roll models, also termed as freezing/tracking  and thawing models(\ Caldwell \& Linder~(2005)). The fast roll models have steep potentials allowing the scalar field to mimick the background matter/radiation, and to remain subdominant for most of the history of the universe.  Only at late times, the field becomes dominant and starts behaving like a component with negative pressure driving  the acceleration of the universe. On the other hand, slow roll or thawing models are similar to the inflaton field, where the potential is sufficiently flat. Due to this, kinetic energy of the field is much smaller than the potential energy and field is initially frozen due to large Hubble damping in the early universe. That is why initially $w\sim -1$, and the energy density of the scalar field is nearly constant and have a negligible contribution to the total energy budget of the universe. But as the  background radiation/matter gets diluted due to the expansion of the universe, the background energy density becomes comparable to the scalar field energy density. At this moment, the scalar field starts thawing and slowly evolves away from the frozen $w\sim -1$ state towards a higher value of $w$. However, in this case some degree of fine tuning is needed in order to get a suitable late time evolution. One of the early studies for thawing models was by Alam et al. (\ Alam et al.~(2003)) where they studied the decaying dark energy models which have thawing type cosmological evolution.
 Recently, Scherrer and Sen (\ Scherrer \& Sen~(2008a))have studied the thawing scalar field models with a nearly flat potential satisfying the slow roll conditions.  They have shown that under such conditions, thawing models with variety of potentials evolve in a similar fashion and results in a unique analytic expression for the equation of state $w(z)$. This result was later extended to the case of phantom models (\ Scherrer \& Sen~(2008b)) and tachyon scalar field models (\ Ali et al.~(2009)). In another work, thawing models ( for both canonical and non canonical scalar field) have been studied without the assumption of the slow roll condition (\ Sen et al.~(2010)).  In a more recent work, Dutta and Scherrer (\ Dutta \& Scherrer~(2011))have studied the slow roll freezing scalar field models and derived an analytical expression for the equation of state $w$ as a function of the energy density of the scalar field $\Omega_{\phi}$.   

There is another elegant way to explain the present cosmic acceleration. This is achieved through the large scale modification of  the gravitational sector of the theory. These models include DGP model (\ Dvali et al.~(2000)), Cardassian model (\ Freese \& Lewis~(2002)), modified GCG model (\ Barreiro \& Sen~(2004)), f(R) theories (\ Starobinsky~(2007),\ Hu \& Sawicki~(2007),Amendola et al.~(2007)), galileon models (\ Nicolis et al.~(2009),\ Chow \& Khoury~(2009)) as well as models having non local gravitational interactions (\ Deser \& Woodard~(2007)).
 
With current obsevational data, the concordance $\Lambda$CDM indeed appears to provide a very good fit to the present observations. Hence, for any given dark energy model (different from $\Lambda$CDM), it is natural to ask how to distinguish it from $\Lambda$CDM. Keeping this in mind, different diagnostic measures have been proposed. $Om$ diagnostic and the Statefinders are two elegant examples for such diagnostic measures. Recently, another interesting diagnostic measure, `{\it Statefinder hierarchy}, $S_{n}$' (with $n \geq 2$), has been introduced by Arabsalmani and Sahni (\ Arabsalmani \& Sahni~(2011)). It was shown that while all the $S_{n}$'s remain fixed at unity during the entire evolution of the universe for concordance $\Lambda$CDM, they are in general time dependent for dark energy models. Hence they can be used not only to compare different dark energy models with $\Lambda$CDM but also to compare different theoretical approaches to explain present cosmic acceleration.

In the present investigation, we consider thawing dark energy models involving canonical scalar fields. We do not assume the slow roll conditions on the potential of the field. We first use the Statefinder diagnostic $S_{n}$ to compare different thawing models with other dark energy models  as well as with modified gravity proposal such as DGP model. Next, we use the latest observational results involving background cosmology to put constraints on the deviations of the thawing models from the $\Lambda$CDM behaviour. We also forecast constraints from future observations. 

\vspace{2mm}
\section{Thawing scalar field models}
We assume that  dark energy is decsribed by a minimally coupled scalar field $\phi$ with canonical kinetic energy, governed by the equation of motion
\begin{equation}
\ddot{\phi} + 3 H \dot{\phi} + \frac{dV}{d\phi} = 0,
\label{sf}
\end{equation}
where $H$ is the Hubble parameter determined by the Einstein equation
\begin{equation}
3 H^{2} = \rho_{m} + \rho_{\phi}.
\label{hub}
\end{equation}
We work in the natural unit $8\pi G = c = 1$. Here, $\rho_{m}$ is the non-relativistic matter  (pressure $ p = 0$) content of the universe and the $\rho_{\phi}$ is the energy density of the scalar field. We assume a spatially flat universe, so that $\Omega_{m} + \Omega_{\phi} = 1$.

\noindent
With this, one can form an autonomous system using equations (\ref{sf}) and (\ref{hub}) (see\ Scherrer \& Sen~(2008a)) for detail derivation):
\begin{eqnarray}
\gamma^{'} &=& -3\gamma(2-\gamma)+ \lambda(2-\gamma)\sqrt{3\gamma\Omega_{\phi}},\\
\Omega^{'}_{\phi} &=& 3(1-\gamma)\Omega_{\phi}(1-\Omega_{\phi}),\\
\lambda^{'} &=& - \sqrt{3}\lambda^{2}(\Gamma-1)\sqrt{\gamma\Omega_{\phi}}.
\label{lam}
\end{eqnarray}
Here $\gamma = 1+ w_{\phi}$ ($w_{\phi}$ being the equation of state for the scalar field), and 
\begin{eqnarray}
\lambda &=& -\frac{1}{V}\frac{dV}{d\phi},\\
\Gamma &\equiv&  V\frac{d^2V}{d\phi^2}/(\frac{dV}{d\phi})^2.
\end{eqnarray}

\begin{figure}
\includegraphics[width=8cm]{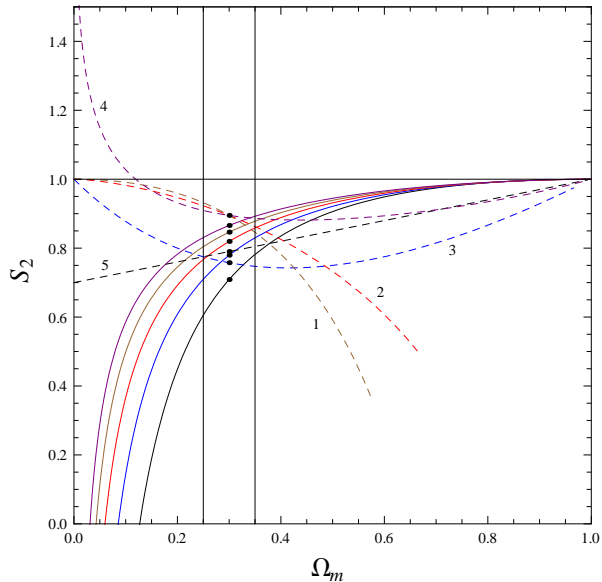}
\caption{The panel shows the Statefinder $S_{2}$ versus $\Omega_{m}\equiv \Omega_{m0}\frac{(1+z)^3}{h^2}$. Larger value of $\Omega_{m}\mapsto1$ corresponds to the distant past ($z>>1$) whereas, the smaller value $\Omega_{m}\mapsto0$ corresponds to the future. $ \Omega_{m0} = 0.3 $ for this plot. The different \textbf{solid lines}  are (Right to Left) : Thawing quintessence model with $\Gamma=0$, $ \Gamma=0.5 $, $ \Gamma=1 $, $ \Gamma=1.5 $, $ \Gamma=2$ respectively. The \textbf{dashed lines} corresponds to CG (1), GCG (2), DGP (3), CPL (4), $ w=-0.8 $ (5). The region between two vertical solid lines roughly corresponds the present epoch.}
\end{figure}

\begin{figure}

\includegraphics[width=8cm]{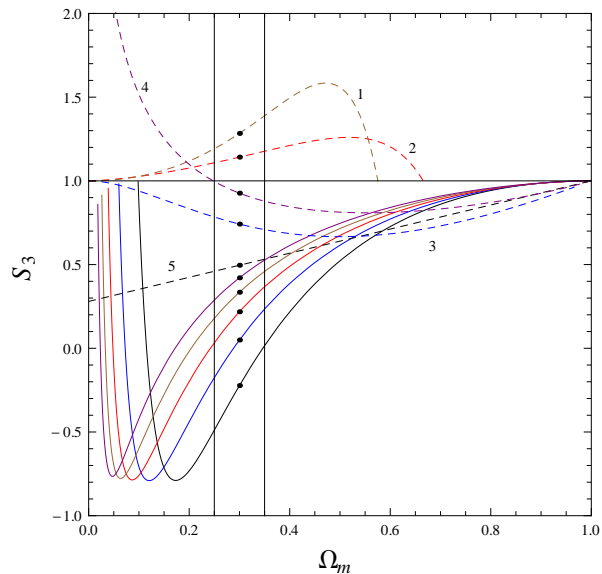}
\caption{Same as Fig.1 but for $S_{3}$.}
\end{figure}

\noindent
Given the initial conditions for $\gamma$, $\Omega_{\phi}$ and $\lambda$, one can solve the above system for any potential  $V(\phi)$ (The information about the form of the potential is encoded in $\Gamma$). As our goal is to study the thawing models, the scalar field is initially frozen due to large Hubble damping (i.e $\dot{\phi}_{i} \sim 0$). Hence $w_{\phi} \sim  -1$ initially which gives $\gamma = 0$ as initial condition. We shall not assume slow-roll conditions for the potential $V(\phi)$. Hence, the initial value $\lambda_{i}$ is a parameter in our model.  This initial value of $\lambda_{i}$ determines the deviation of $w_{\phi}$ from the initial $w_{\phi} \sim -1$  frozen state as the universe evolves. For smaller $\lambda_{i}$, the deviation is negligibly small, and the scalar field behaves as a cosmological constant at all time. For larger values of $\lambda_{i}$, $w_{\phi}$ deviates substantially from $-1$ as the universe evolves.  In general, the contribution of scalar field to the total energy density  is negligible in the early universe. Nevertheless one has to fine tune the initial value of $\Omega_{\phi}$ in order to have its correct contribution at present. Hence $\Omega_{\phi}(initial)$ is related to the $\Omega_{\phi} (z = 0)$. With this we evolve our system from the intial era $z=1000$ till any redshift we want. We consider various power-law potentials for which $\Gamma$ is a constant. In recent past, Clemson and Liddle have considered thawing models using a multiparameter extension of the exponential potential (\ Clemson \& Liddle~(2009)).

\vspace{2mm}
\section{Statefinder Hierarchy and Parameter Phase Space}
One can Taylor expand the scale factor of the universe around the present era ($t=t_{0}$) as follows:
\begin{equation}
a(t) = a(t_{0}) + a(t_{0})\sum^{\infty}_{n=1} \frac{\alpha_{n}(t_{0})}{n!}\left[H_{0}(t-t_{0})\right]^n,
\label{state}
\end{equation} 
where,
\begin{equation}
\alpha_{n} = \frac{d^{n}a}{dt^n}/(aH^{n}).
\label{alpha}
\end{equation}
 It is easy to check that $-\alpha_{2} \equiv q$  is the deceleration parameter. Similarly $\alpha_{3}$ is related to the Statefinder $r$ or jerk $j$ and $\alpha_{4}$ is related to the snap $s$ and so on. Using the equations (\ref{state}) and (\ref{alpha}), one can now define the Statefinder Hierarchy $S_{n}$ (\ Arabsalmani \& Sahni~(2011)) as:
 \begin{eqnarray}
 S_{2} &=& \alpha_{2} + \frac{3}{2}\Omega_{m}\\
 S_{3} &=& \alpha_{3}\\
 S_{4} &=& \alpha_{4} + \frac{3^2}{2}\Omega_{m}\\
 S_{5} &=& \alpha_{5} - 3 \Omega_{m} - \frac{3^3}{2}\Omega^{2}_{m}\\
 S_{6} &=& \alpha_{6} + \frac{3^3}{2}\Omega_{m} + {3^4}\Omega^{2}_{m} + \frac{3^4}{4}\Omega^{3}_{m}  \hspace{2mm} \hspace{2mm} and \hspace{2mm} so \hspace{2mm} on.
 \end{eqnarray}

\noindent
It is remarkable to see that for $\Lambda$CDM, $S_{n} = 1$ during the entire history of the universe. In the rest of this section we use different combinations of $S_{n}$, to study the evolution of thawing models and compare them with other popular dark energy and modified gravity models.We consider the following models:
\begin{itemize}
\item{Dark Energy with a constant equation of state, $w = w_{0}= constant$.}
\item{Dark Energy with a CPL parametrized equation of state (\ Chevallier \& Polarski~(2001), \ Linder~(2003)), $ w = w_{0} + w_{a}(1-a)$, where $w_{0}$ and $w_{a}$ are constants.}
\item{Dark Energy with a Generalized Chaplygin Gas (\ Bento et al.~(2002)) type equation of state, $w = -\frac{A_{s}}{A_{s}+(1-A_{s})a^{-3(1+\beta)}}$, where $A_{s}$ and $\beta$ are constant. $\beta =1$ gives the standard Chaplygin gas.}  
\item{Dvali, Gabadadze and Porrati (DGP) model (\ Dvali et al.~(2000)).}
\end{itemize}

\begin{figure}
\includegraphics[width=8cm]{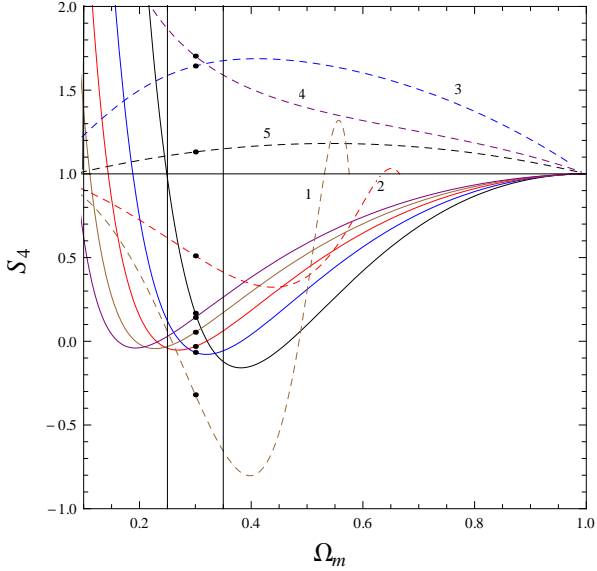}
\caption{Same as Fig.1 but for $S_{4}$.}
\end{figure}

In Fig.1, we show the behaviours of the models in the $S_{2}-\Omega_{m}$ phase plane.  The figure shows that in this phase plane which involves the second derivative of the scale factor, the models are highly degenerate around the present value of $\Omega_{m}$.  As we go to the higher order $S_{n}$'s, one can remove these degeneracies among various models. This is revealed in the subsequent figures, Fig. 2 and Fig. 3 where we show the phase portrait in the $S_{3}-\Omega_{m}$ and $S_{4}-\Omega_{m}$ planes respectively.  In Fig. 4 we  show the phase portrait in $S_{4}-S_{3}$ plane where again the degeneracies among different models are broken. We should comment on $S_{3}$ in particular.  From Fig. 2 and Fig. 4, one can see that $S_{3}$ can be used to break the degeneracies among the thawing models themselves as its present day value  varies substantially for different thawing potentials. Similarly, $S_{4}$ lets us differentiate between thawing and other models. We should mention that although we assume $\Omega_{m0} = 0.3$ for these figures, the conclusions remain same for other values of $\Omega_{m0}$.

\begin{figure}
\includegraphics[width=8cm]{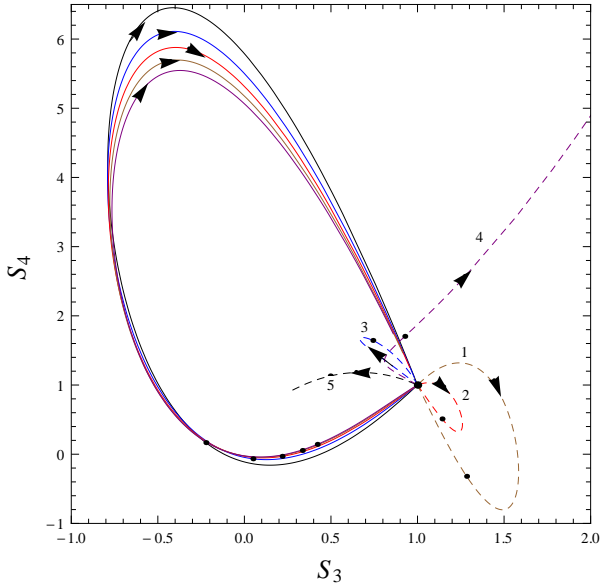}
\caption{The phase plane for $ S_{4} vs. S_{3}$. Dashed lines are same as Fig 1. The solid lines (top to bottom): Thawing quintessence model with $\Gamma=0$, $ \Gamma=0.5 $, $ \Gamma=1 $, $ \Gamma=1.5 $, $ \Gamma=2$ respectively. $\Omega_{m0} = 0.3$ for this plot. The point (1,1) is $ \Lambda$CDM.}
\end{figure}

\begin{figure}
\includegraphics[width=8cm]{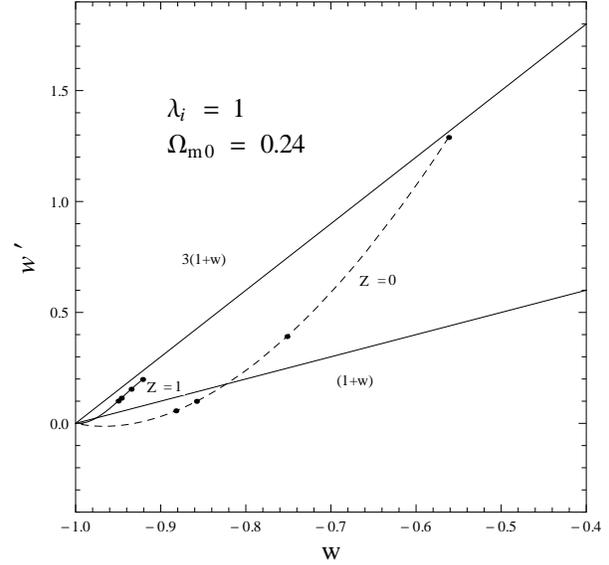}
\caption{The panel shows the dynamics of scalar field in $w-w^{\prime}$ plane at redshifts $z = 0$ \textit{(solid)}and $z = 1$ \textit{(dashed)}.The dots \textit{(black)} represents the potentials $V = \phi, \phi^{2},\phi^{-2},\phi^{-1} $ (from top to bottom, for each z)respectively.}
\end{figure} 
With future data, one can probe the background expansion at higher redshifts thereby probing the higher derivatives of the scale factor. Hence Statefinder Hierarchy will   be quite useful to distinguish various dark energy models.  
We further study the $w^{\prime}-w$ phase plane for different thawing dark energy models. Here `prime' denotes the differentiation with respect to $\log a$ . As discussed in the previous section, for smaller values of $\lambda_{i}$, all the thawing models behave similar to the cosmological constant. So to study the $w^{\prime}-w$ phase plane, we assume $\lambda_{i} =1$ for which the thawing models behave significantly different from cosmological constant.    Linder and Caldwell have previously showed that the thawing models are constrained to lie between $(1+w) \leq w^{\prime} \leq 3(1+w)$ region in $w^{\prime}-w$ phase plane.  In Fig. 5 and Fig. 6, we show different thawing models in the $w^{\prime}-w$ phase plane at different redshifts and compare them with the constraint proposed by Linder and Caldwell (\ Caldwell \& Linder~(2005)) (LC). It can be seen that for higher redshifts, all the thawing models are inside the LC bound. This is obvious as all the thawing models behave very close to cosmological constant for larger redshifts and hence satisfy the LC bound. But at the present epoch (z=0), for smaller values of $\Omega_{m0}$, thawing models with certain  potentials ( as shown in Figure 5) are outside the LC bound. But as one increases the value of $\Omega_{m0}$ (see Figure 6), they tend to satisfy the LC bound. Hence thawing models which deviate significantly from the $\Lambda$CDM behaviour at smaller redshift tend to satisfy the LC bound for higher values of $\Omega_{m0}$ whereas for lower values of $\Omega_{m0}$, some of the potentials do not satisfy the LC bound.

\begin{figure}
\includegraphics[width=8cm]{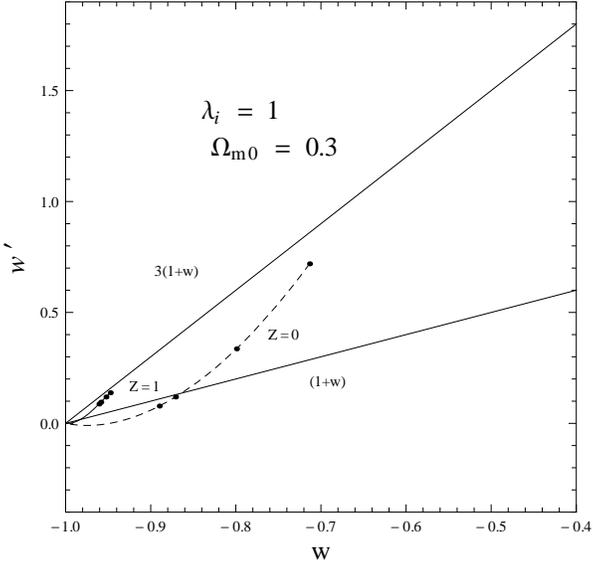}
\caption{Same as Figure 5 but with $\Omega_{m0} = 0.3$}
\end{figure}

\section{Observational Constraints}

In this section, we constrain the parameters in the thawing scalar field models with the assumption of a flat Universe by using the latest observational data including the Type-1a Supernovae Union2 compilation (\ Amanullah et al.~(2010)), the BAO (\ Eisenstein et al.~(2005)) measurement from the SDSS (\ Percival et al.~(2007),\ Percival et al.~(2010)), the CMBR measurement given by WMAP7 (\ Komatsu et al.~(2011)) observations, the H(z) data from HST key Project (\ Freedman et al.~(2001)) and a simulated dataset (\ Holsclaw et al.~(2010)) based on the upcoming  JDEM SN-survey containing around 2300  Type-1a Supernovae.

\subsection*
{Type Ia Supernovae}
We consider the Supernovae Type Ia observation (\ Amanullah et al.~(2010) which is one of the direct probes for the late time acceleration. It measures the apparent brightness of the Supernovae as observed by us which is related to the luminosity distance $d_{L}(z)$ 
defined  as 
\begin{equation}
d_{L}(z) = (1+z)\int_0^z\frac{dz^{\prime}}{H(z^{\prime})}\end{equation}

With this, we construct the distance modulus $\mu$ which is experimentally measured
\begin{equation}
\mu = m-M = 5\log\frac{d_{L}}{Mpc}+25
\end{equation}
Where m and M are the apparent and absolute magnitudes of the Supernovae which are logarithmic measure of flux and luminosity respectively.

\begin{figure}
\psfrag{lambdai}{$\lambda_{i}$}
\psfrag{omegam}{$\Omega_{m0}$}
\includegraphics[width=8cm]{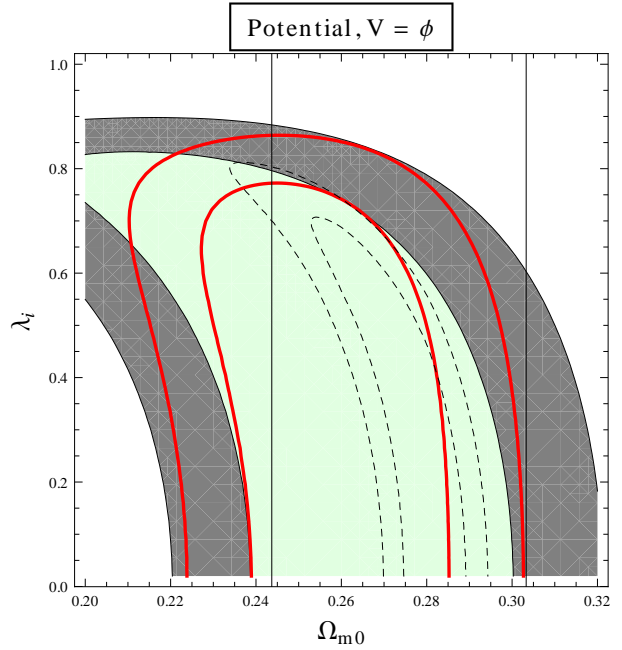}
\caption{The 1$\sigma$ and  2$\sigma$ confidence contours in $\lambda$i - $\Omega_{m0}$ plane for $ \Gamma = 0 $.The shaded regions are constraints from SN+BAO data while the thick lines are constraints from SN+BAO+CMB+H(z) data respectively. The dashed lines are for the simulated JDEM data. The two vertical lines represent the WMAP7 bound on $\Omega_{m0}$}

\end{figure}

\begin{figure}
\psfrag{lambdai}{$\lambda_{i}$}
\psfrag{omegam}{$\Omega_{m0}$}
\includegraphics[width=8cm]{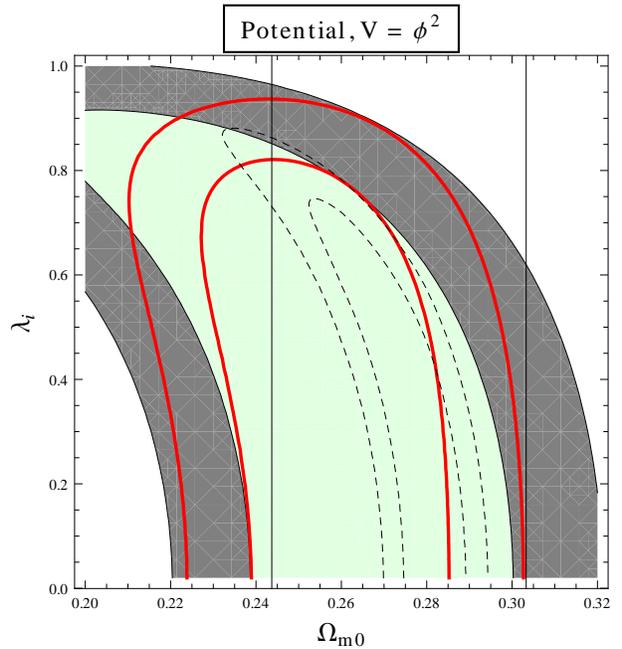}
\caption{Same as Figure  7, but with $ \Gamma = 0.5 $.}

\end{figure}

\subsection*
{Baryon Acoustic Oscillation}
Another observational probe that has been widely used in recent times to constrain dark energy models is related to the data from the BAO distance measurements obtained at $z = 0.2$ and $z = 0.35$ from the joint analysis of the 2dFGRS and SDSS data (\ Eisenstein et al.~(2005),\ Percival et al.~(2007),\ Percival et al.~(2010)).
In this case,one needs to calculate the parameter $D_{v}$ which is related to the angular diameter distance as follows
\begin{equation}
D_{v} = \left[\frac{z_{BAO}}{H(z_{BAO})}\left(\int_0^{z_{BAO}}\frac{dz}{H(z)}\right)^2\right] ^{1/3}
\end{equation}
For BAO measurements we calculate the ratio 

{\Large$\frac{D_{v}(z = 0.35)}{D_{v}(z = 0.20)} $}
 \vspace{1mm}
this ratio is a relatively model independent quantity and has a value $1.736 \pm 0.065$.

\subsection*
{Cosmic Microwave Background}
The CMB is sensitive to the distance to the decoupling epoch via the locations of peaks and troughs of the acoustic oscillations. We employ the ``WMAP distance priors''  given by the five-year WMAP observations.This includes the``acoustic scale'' $l_{A}$,the `` shift parameter '' R and the redshift of the decoupling epoch of photons $z_{*}$
The acoustic scale $l_{A}$ describes the distance ratio $\frac{D_{A}(z_{*})}{r_{s}(z_{*})}$
\begin{equation}
l_{A} \equiv (1+z_{*})\frac{\pi D_{A}(z_{*})}{r_{s}(z_{*})}
\end{equation}
where $(1+z_{*})$ factor arises because $D_{A}(z_{*})$ is the proper angular diameter distance, where $r_{s}(z_{*})$ is the comoving sound horizon at $z_{*}$

We use the fitting function of $z_{*}$ proposed by Hu and Sugiyama (\ Hu \& Sugiyama~(1996)) 
\begin{equation}
z_{*} = 1048[1+0.00124(\Omega_{b}h^2)^{-0.738}][1+g_{1}(\Omega_{m}h^2)^{g_{2}}]
\end{equation}
\begin{equation}
g_{1} = \frac{0.0783(\Omega_{b}h^2)^{-0.238}}{1+39.5(\Omega_{b}h^2)^{0.763}} , 
 g_{2} = \frac{0.560}{1+21.1(\Omega_{b}h^2)^{1.81}}
\end{equation}

The shift parameter R (\ Komatsu et al.~(2011)) is responsible for the distance ratio $\frac{D_{A}(z_{*})}{H^{-1}(z_{*})}$ given by
\begin{equation}
R(z_{*}) \equiv \sqrt{\Omega_{m0}H_{0}^{2}}(1+z_{*})D_{A}(z_{*})
\end{equation}
The constraint on the shift parameter $R(z_{*})$ from the WMAP observations is quoted as $R(z_{*}) = 1.715\pm 0.021$.

\subsection*
{ Expansion history H(z) data}
We use new determinations of the cosmic expansion history from red-envelope galaxies. Stern et al. (\ Stern et al.~(2010)) have obtained a high-quality spectra with the Keck-LRIS spectrograph of red-envelope galaxies in 24 galaxy clusters in the redshift range $0.2< z< 1.0$. They complemented these Keck spectra with high-quality, publicly available archival spectra from the SPICES and VVDS surveys.With this, they presented 12 measurements of the Hubble parameter H(z) at different redshift. The measurement at z = 0 is from HST Key project \citep{Freedman2001}.

\begin{figure}
\psfrag{lambdai}{$\lambda_{i}$}
\psfrag{omegam}{$\Omega_{m0}$}
\includegraphics[width=8cm]{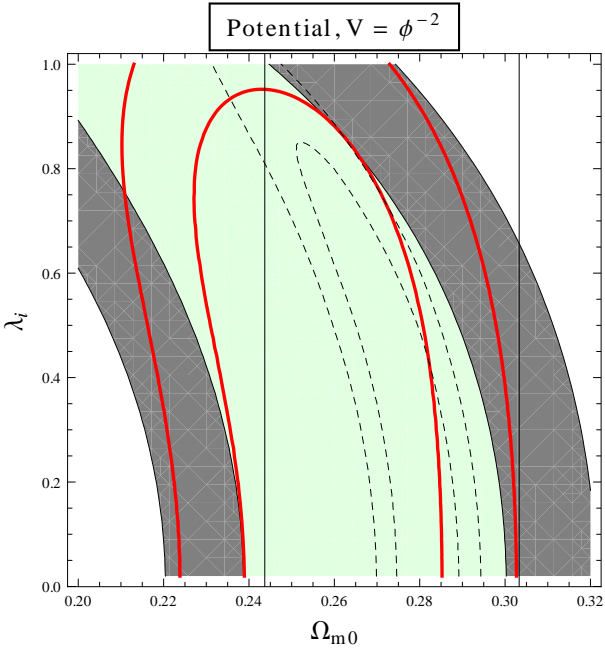}
\caption{Same as Figure 7 but with $ \Gamma = 1.5 $.}

\end{figure}

\begin{figure}
\psfrag{lambdai}{$\lambda_{i}$}
\psfrag{omegam}{$\Omega_{m0}$}
\includegraphics[width=8cm]{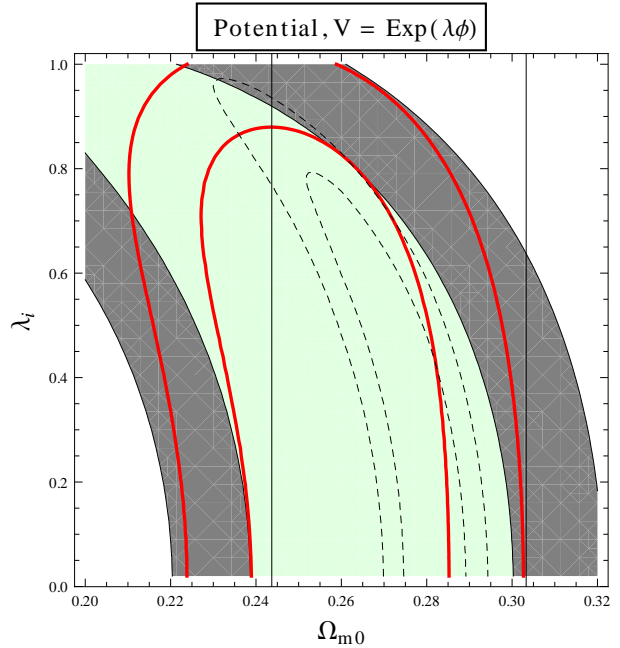}
\caption{Same as Figure 7 but with $\Gamma = 1$.}
\end{figure}

\subsection*
{Simulated SN data}

To look at what can be achieved in future, we also use a simulated dataset (\ Holsclaw et al.~(2010)) based on the upcoming  JDEM SN-survey containing around 2300 SNe. These are distributed over a redshift range from $z=0$ to $z=1.7$. We assume a simplified error model where the errors are  same for all supernovae and independent of redshift. We assume a statistical error of $\sigma = 0.13$ magnitude, as expected from JDEM-like future surveys(\ Aldering et al.~(2004)) . We use a $\Lambda$CDM model with $H_{0} = 72$ km/s/Mpc and $\Omega_{m0} = 0.27$ to simulate the data.

\begin{table*}
\begin{tabular}{|c|c|c|}
\hline
\\
\includegraphics[width=5.5cm,height=5cm]{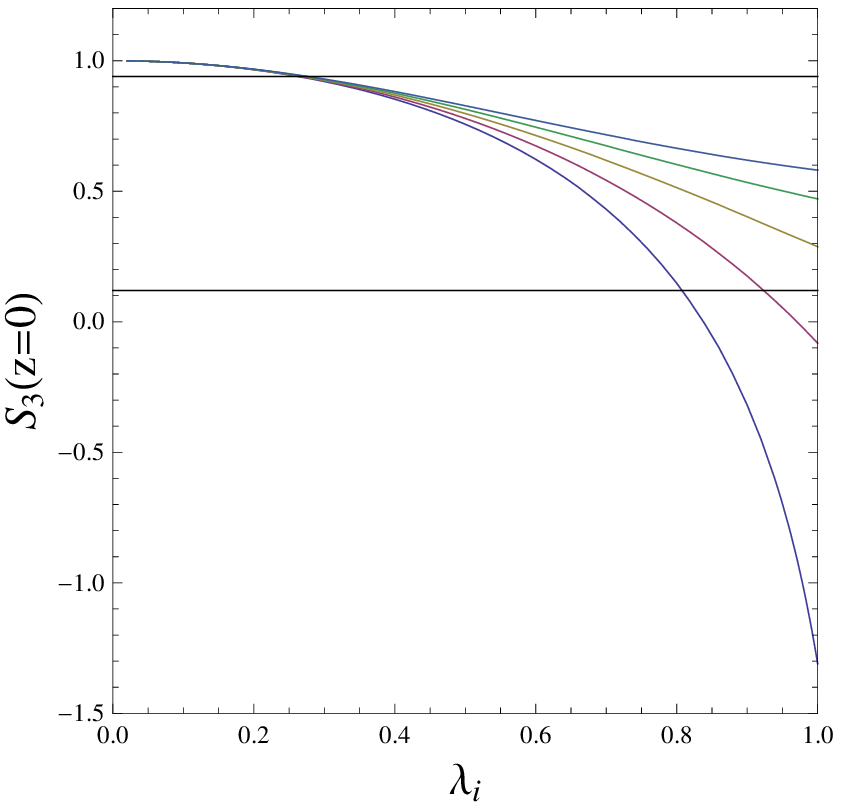} 
\includegraphics[width=5.5cm,height=5cm ]{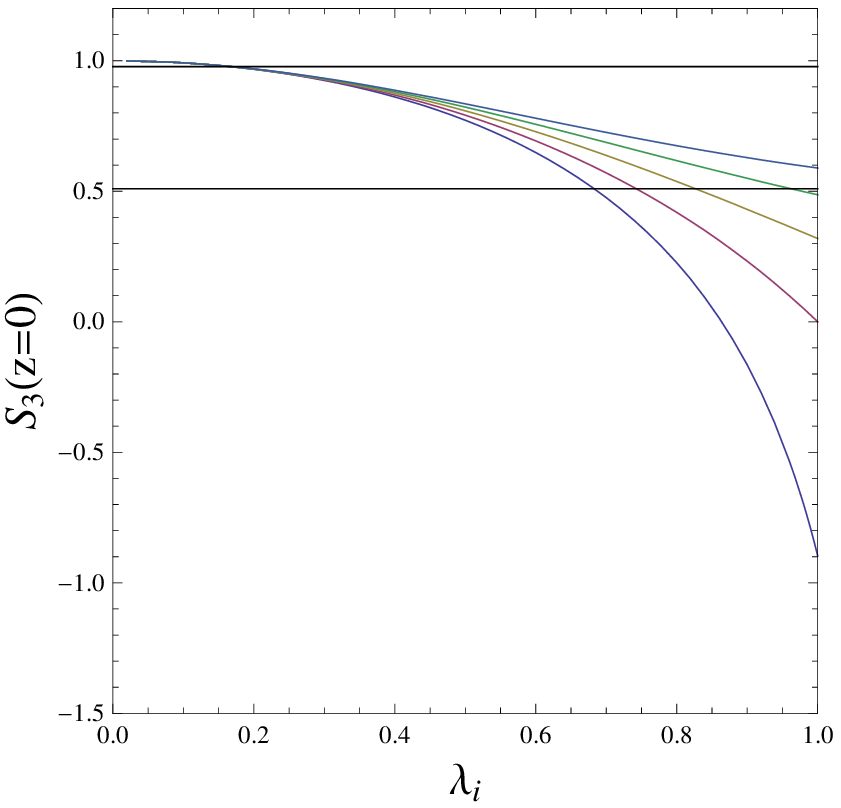} 
\includegraphics[width=5.5cm, height=5cm]{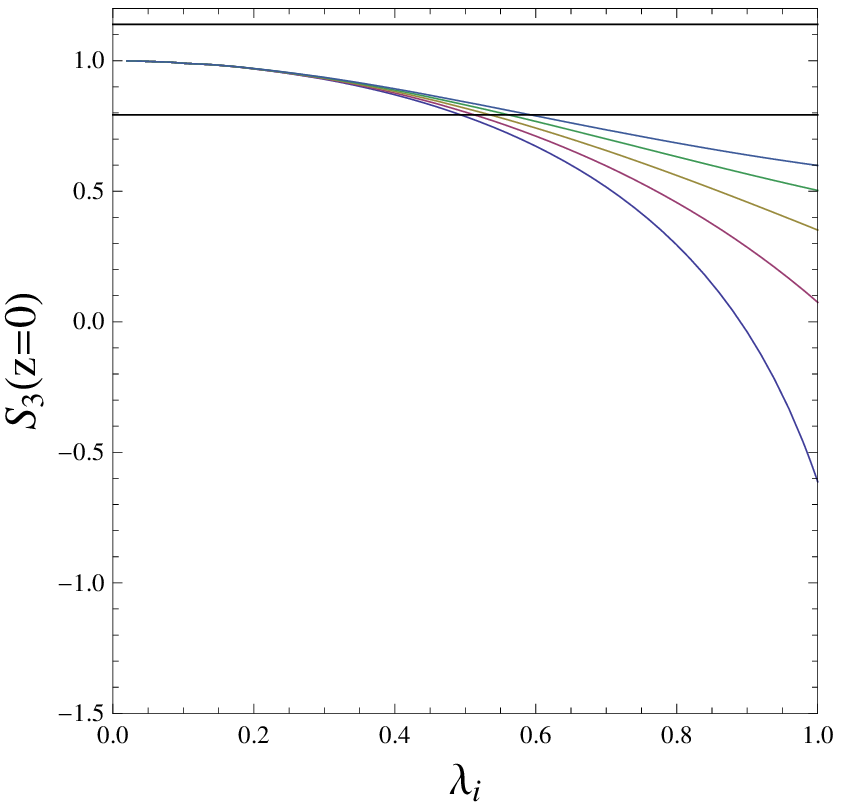} 
\\
\hline
\end{tabular}

\vspace{2mm}
{{\bf Figure 11}: The behaviour of $S_{3}$ at present ($z=0$) as a funtion of $\lambda_{i}$ for different thawing potentials. The lines represent $\Gamma = 0, 0.5, 1, 1.5, 2$ from bottom to top. $\Omega_{m0} = 0.24,0.26,0.28$ (from left to right) respectively. The horizontal lines represent the bound on $S_{3}$ at $z=0$ due to the JDEM simulated data.}
\end{table*}

\subsection*{Constraints on Model Parameters}

Thawing models are described by three parameters, namely $\lambda_{i}$, $\Gamma $ and $ \Omega_{m0} $. As mentioned in the previous sections, $\lambda_{i}$ describes the deviation of the thawing model from $\Lambda$CDM ($\lambda_{i} = 0$ represents the $\Lambda$CDM model), and $\Gamma$ encodes the information about the potential. Hence constraining these two parameters essentially constrain the deviation of different thawing models from $\Lambda$CDM.  In Figures 7, 8, 9 and 10, we show the constrained region in $\lambda_{i}-\Omega_{m0}$ plane for different potentials. In doing this, we use the data coming from different recent observations e.g SnIa, BAO, CMB as well as H(Z) measurements. We also use the simulated data for the upcoming SN-surveys like JDEM to constrain the parameter space.  In these figures, we also show the WMAP7 bound on $\Omega_{m0}$.

From the figures it is evident that when we consider the current data, there is always an upper bound on $\lambda_{i}$. This upper bound is most prominent for the linear potential. Hence this gives strong constraint on deviation from the $\Lambda$CDM behaviour for the linear potential. For Exponential or Inverse Squared potential, this deviation is not strongly constrained and $\lambda_{i}$ as large as 1 is allowed even at $2-\sigma$ confidence level. Moreover if one takes into account the WMAP7 bound on $\Omega_{m0}$, current data always allow the $\Lambda$CDM behaviour ($\lambda_{i} =0$) for all the potentials. 

But this changes as one does similar analysis with the simulated data for future SN-surveys like JDEM (\ Holsclaw~(2010)). With this, considering the WMAP7 bound on $\Omega_{m0}$, although $\Lambda$CDM ($\lambda_{i} =0$) is allowed for some values of $\Omega_{m0}$, for other values of $\Omega_{m0}$, $\lambda_{i}$  has a lower bound different from zero. This clearly rules out $\Lambda$CDM even at $2-\sigma$ confidence level. For example, consider the constraints obtained from simulated JDEM dataset in case of linear potential (Fig.7). The contours are shown by the dashed line. For $\Omega_{m0}=0.25$ (which is allowed by the WMAP7 constraint on $\Omega_{m0}$), the lower bound on $\lambda_{i}$ is 0.75 at $2-\sigma$ confidence level. Similarly, in case of an exponential potential (Fig.10), the corresponding lower bound on $\lambda_{i}$ for the same $\Omega_{m0}$ is 0.85 at $2-\sigma$ confidence level. This rules out the $\Lambda$CDM even at $2-\sigma$ confidence level. For all the potentials that we considered, one can rule out $\Lambda$CDM at $2-\sigma$ confidence level for some values of $\Omega_{m0}$ using simulated data for future-SN survey like JDEM. Although we study a generic class of dark energy, e.g the thawing models, this shows the advantage of the large scale surveys like JDEM to distinguish different dark energy models from $\Lambda$CDM which is difficult with the presently available data.

Next we study how effective is the simulated data in distinguishing different dark energy models using the {\it Statefinder Hierarchies}. In section 2, we already mentioned that $S_{3}$ is the most promising parameter to distinguish different dark energy models.  Keeping this in mind, we calculate the bound on the present day value of $S_{3}$, using the simulated JDEM data for different values of $\Omega_{m0}$. In figure 11, the regions inside the horizontal lines show this bound at $2-\sigma$ confidence level. This bound on $S_{3}$ is valid for all thawing models  with power law potentials and is independent of the power law form. As one can see, the bound becomes narrower as one increases the $\Omega_{m0}$. Moreover, it is clear that as one decreases the value of $\Omega_{m0}$, e.g. $\Omega_{m0} = 0.24$, $S_{3}(z=0)=1$ is outside the $2-\sigma$ allowed region ruling out the $\Lambda$CDM. But for higher values of $\Omega_{m0}$, $\Lambda$CDM is allowed. This is consisnent with our earlier conclusions from Fig 7 - Fig 10.

In the same figure, we also show the behaviour of $S_{3} (z=0)$ as a function of $\lambda_{i}$ for different potentials. 
For $\Omega_{m0} = 0.24$, although $\Lambda$CDM is ruled out, but all the thawing models more or less are inside the allowed region of $S_{3}$ making them difficult to distinguish.  As one increases the $\Omega_{m0}$, models with higher $\lambda_{i}$ start going out of the allowed region of $S_{3}$. This helps to distinguish different thawing models with higher $\lambda_{i}$. We should emphasize the fact that for smaller $\lambda_{i}$, all the thawing models have exactly similar behaviour which is same as $\Lambda$CDM. It is only for higher $\lambda_{i}$, the different thawing models behaves differently.

\section{Discussion}

We look at observational constraints on the thawing scalar field models which have been proposed to explain late time cosmic acceleration. We first study the degeneracies of different thawing models with other popular dark energy models using the recently proposed ` Statefinder Hierarchy $S_{n}$ '.  We show that as one starts probing the higher order $S_{n}$, different thawing models not only become distinguishable among themselves, they also behave significantly different from other dark energy models including $\Lambda$CDM. With expected future data at significantly higher redshifts, this can be a smoking gun to distinguish different dark energy models.  We also study the behaviour of different thawing models in the $w^{\prime}-w$ phase plane and compare those with the previously obtained LC bound. Using various observational data related to background evolution of the universe, we constrain the deviation of the thawing model from $\Lambda$CDM. With current precision level of the observational data ( at least for the background evolution), the models cannot be distinguished from the $\Lambda$CDM. But with simulated data for large future surveys like JDEM,  we show that for some values of $\Omega_{m0}$, the data may distinguish thawing models with $\Lambda$CDM. Moreover, with future data, the $S_{3}$ parameter may be quite useful in distinguishing different thawing models among themselves for certain plausible values of $\Omega_{m0}$.

As a possible extension of this work, it will be really interesting to do a similar study with freezing type of scalar fields. As we mention in the introduction, Dutta and Scherrer (\ Dutta \& Scherrer~(2011)) have recently studied the slow-roll freezing scalar field models. In such a setup, one can calculate the `Statefinder Hierarchy' and compare them with those for the thawing models. With future high precision data, such study can be useful to check which kind of scalar field model is actually preferred by the data. This will be our future goal.

\section{Acknowledgement}
AAS acknowledges the financial support provided by the SERC, DST, Govt. of India, through major research project grant (Grant No: SR/S2/Hep-43(2009)) and the hospitality provided by IUCAA, Pune where part of this work has been done. GG acknowledges the Senior Research fellowship provided by CSIR, Govt. of India and the hospitality provided by TIFR, Mumbai where part of this work has been done. AAS and GG gratefully acknowledge the suggestions and comments from V. Sahni during the earlier stages of the work.

\end{document}